\documentclass[useAMS,usenatbib]{mn2e}
\usepackage{graphicx,subfigure,ram,natbib}

\usepackage[T1]{fontenc}

\begin{document}

\title[Pc-scale jet of 3C 286]{Parsec-scale jet properties of the gamma-ray quasar 3C~286}

\author[An et al.]{
T. An$^{1,2}$\thanks{Email: antao@shao.ac.cn}, B.-Q. Lao$^{1}$, W.~Zhao$^{1}$, P. Mohan$^{1}$, X.-P. Cheng$^{1,3}$, Y.-Z. Cui$^{1}$, Z.-L. Zhang$^{1}$ \\
$^{1}$ Shanghai Astronomical Observatory, Chinese Academy of Sciences, 200030 Shanghai, China \\
$^{2}$ Key Laboratory of Radio Astronomy, Chinese Academy of Sciences, 210008 Nanjing, China \\
$^{3}$ University of Chinese Academy of Sciences, 19A Yuquanlu, Beijing 100049, China
}

\date{}

\maketitle

\begin{abstract}
The quasar 3C~286 is one of two compact steep spectrum sources detected by the {\it Fermi}/LAT. Here, we investigate the radio properties of the parsec(pc)-scale jet and its (possible) association with the $\gamma$-ray emission in 3C~286. The Very Long Baseline Interferometry (VLBI) images at various frequencies reveal a one-sided core--jet structure extending to the southwest at a projected distance of $\sim$1 kpc. The component at the jet base showing an inverted spectrum is identified as the core, with a mean brightness temperature of $2.8\times 10^{9}$~K.  The jet bends at about 600 pc (in projection) away from the core, from a position angle of $-135^\circ$ to $-115^\circ$. Based on the available VLBI data, we inferred the proper motion speed of the inner jet as $0.013 \pm 0.011$ mas yr$^{-1}$ ($\beta_{\rm app} = 0.6 \pm 0.5$), corresponding to a jet speed of about $0.5\,c$ at an inclination angle of $48^\circ$ between the jet and the line of sight of the observer. The brightness temperature, jet speed and Lorentz factor are much lower than those of $\gamma$-ray-emitting blazars,
 implying that the pc-scale jet in 3C~286 is mildly relativistic. Unlike blazars in which $\gamma$-ray emission is in general thought to originate from the beamed innermost jet, the location and mechanism of $\gamma$-ray emission in 3C~286 may be different as indicated by the current radio data. Multi-band spectrum fitting may offer a complementary diagnostic clue of the $\gamma$-ray production mechanism in this source.
\end{abstract}


%________________________________________________ sections below
%

\section{Introduction}
\label{sec:introduction}

An important observational result of the Energetic Gamma Ray Experiment Telescope (EGRET) on the {\it Compton Gamma-Ray Observatory} \cite[][]{fic94b} is the discovery that the majority of the detected extragalactic $\gamma$-ray sources are identified as blazars, consisting of BL Lacertae objects and flat-spectrum radio quasars \cite[][]{har99}. This was confirmed by the Large Area Telescope (LAT) on the {\it Fermi Gamma-ray Space Telescope} which found that active galactic nuclei (AGN) occupy a vast fraction of the {\it Fermi}-detected sources \cite[][]{Abdo10-1FGL} with 98\% of $\gamma$-ray AGN being identified as blazars \cite[][]{ace15,ack15}. A persistent short-timescale variability (from few hundreds of seconds to days) of $\gamma$-ray flux, especially during a flaring state indicates strongly beamed emission originating from a compact region \cite[e.g.][]{2007ApJ...664L..71A,Abdo09}. The $\gamma$-ray emission in blazars is believed to originate in the synchrotron self-Compton process where lower-energy photons from a highly relativistic electron population are up-scattered by the same population, and in the external Compton process where external seed photons from the accretion disk, broad line region clouds or the torus are up-scattered by the relativistic electron population at the jet base. Their discernible signatures can be inferred by modeling the AGN spectral energy distribution \cite[e.g.][]{Pian02,LVT05,AbdoAck10} or from other observables such as the relationship between photon energy density and the Doppler factor \cite[e.g.][]{2006ApJ...646....8F}. 

Among the 1591 {\it Fermi}-detected AGN \cite[][]{ack15}, thirty-two are non-blazar AGN, including two compact steep spectrum (CSS) sources, fourteen radio galaxies, five  narrow-line Seyfert 1 galaxies and six other type AGN \citep[recently reviewed by][]{2016A&ARv..24....2M}. The CSS and GHz-Peaked-Spectrum (GPS) sources constitute a significant fraction of compact radio-loud AGN in flux density limited radio surveys \cite[e.g.][]{GP09} and are characterized by a steep spectrum and compact source size \cite[CSS $<$15 kpc, and GPS $<$1 kpc:][]{fanti85,fanti09}. The source 3C 286 is one of two identified unique $\gamma$-ray emitting CSS sources. The other one is 4C $+$39.23B, but its $\gamma$-ray source identification is suspicious since the $\gamma$-ray emission could arise from the nearby blazar 4C $+$39.23 \citep{2016A&ARv..24....2M}. A detailed study of the jet properties (kinematics, morphology, Doppler boosting factor) in 3C 286 will be helpful for understanding the physical origin and location of the $\gamma$-ray emission from this compact misaligned AGN. By misaligned, we refer to low power AGNs with weakly relativistic jets owing to larger inclination angles towards the observer's line of sight resulting in low Doppler boosting, interpreted in terms of the AGN orientation based unification scheme \cite[e.g.][]{UP95}.

\begin{figure}
\centerline{\includegraphics[width=0.5\textwidth]{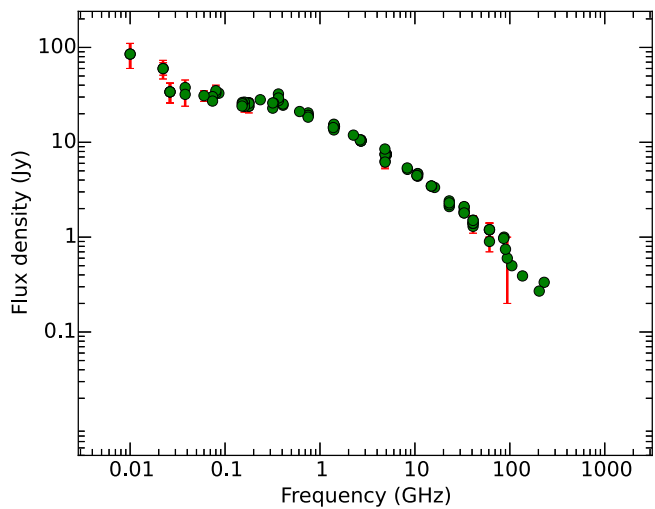}}
\caption{Radio spectrum of 3C~286. We only used the data observed with single dish radio telescopes or connected element interferometers, which reflect the flux density of the entire source. References for data points: (10, 22.25, 26.3, 38, 86, 178, 750) MHz and (2.7, 5.0, 10.7, 14.9) GHz \citep{1980MNRAS.190..903L}; 60 MHz \citep{1968Afz.....4..129A}; 74 MHz \citep{2007AJ....134.1245C}; (80, 160, 178, 318, 365, 408, 750) MHz and (1.4, 2.7, 5.0) GHz \citep{1981A&AS...45..367K}; (80, 160) MHz \citep{1995AuJPh..48..143S}; 151 MHz \citep{1988MNRAS.234..919H,1996MNRAS.282..779W}; (235, 325, 610) MHz \citep{2004ApJ...612..974C}; 1.4 GHz \citep{1992ApJS...79..331W,2010ApJ...714.1170M,2011ApJ...728L..14B}; (2.6, 8.0) GHz \citep{2006ApJS..165..439R}; (2.7, 10.45) GHz \citep{2007A&A...470...83G}; 4.85 GHz \citep{1991ApJS...75.1011G}; (4.85, 8.35, 10.45) GHz \citep{2009A&A...502...61M}; 15 GHz \citep{2011ApJS..194...29R}; (23, 33, 41) GHz \citep{2003ApJS..148...97B}; (23, 33, 41, 61) GHz \citep{2009ApJS..180..283W}; (23, 33, 41, 61) GHz \citep{2009MNRAS.392..733M}; (23, 33, 41)  GHz \citep{2011ApJS..192...15G}; 86 GHz \citep{2010ApJS..189....1A}; (86.9, 105.4, 136.2) GHz \citep{2004MNRAS.352..563C}; (90, 230) GHz \citep{1995A&AS..113..409S}; 94 GHz \citep{2009ApJ...694..222C}.}
\label{fig:spec}
\end{figure}

The quasar 3C~286 \cite[B1328+307, $m=17.3$, $z=0.849$, ][]{coh77} is identified as a CSS quasar \cite[][]{PW82}. The radio spectrum is constructed with data available from the NED\footnote{http://ned.ipac.caltech.edu/}. The shape indicates a steep radio spectrum (Figure~\ref{fig:spec}) with a spectral index of $\alpha = -0.61$ ($S_\nu \propto \nu^\alpha$) between 1.4 and 50 GHz with a turnover at about 300 MHz, below which it is flat till $\sim$ 75 MHz. The source is slightly resolved in the 1.7 GHz Very Large Array (VLA) image, displaying a primary core and a secondary lobe $\sim$ 2.6\arcsec{} (19.4 kpc) to the southwest \cite[][]{spe89}. The radio flux density of the source is rather stable during a multi-band radio monitoring study between 1994 and 1999 spanning 5.5 years \cite[][]{Peng00}. It is thus often used as a flux density calibrator owing to this high and stable radio flux density, and compact size \cite[][]{Per13}. At even higher resolution and higher frequency of 8.4 and 22.5 GHz, the source is resolved into a core--double-lobe structure with a bright central component and two lobes at 0.7\arcsec{} to the east and at 2.6\arcsec{} to the southwest \cite[][]{An04}, with an observed knotty morphology between the core and southwestern lobe. In addition, high polarization has been detected in this source at radio wavelengths \cite[][]{AG95,lud98}, indicating typical properties characterising a CSS quasar. The {\it Hubble Space Telescope} (HST) image shows a compact bright nucleus associated with the radio core \cite[][]{dev97}. A secondary fainter optical feature is seen at $\sim$ 10\arcsec{} east of the quasar, roughly aligning with the eastern lobe and is likely a lightened star forming region impacted by the expanding (relic) radio jet or a distorted nucleus of the merging companion. Similar observations have also been obtained in another well studied CSS quasar, 3C 48 \cite[][]{cha99,Feng05,sto07,An10}.

At pc scales, the Very Long Baseline Interferometry (VLBI) images of 3C~286 display a complex elongated structure with two embedded compact components that can be interpreted as either core--jet or double hot spots \cite[][]{zhang94,jiang96,cotton97,kel98}. The VLBI structure extends toward the southwest to a projected distance of 700 pc \cite[94 mas,][]{zhang94}. High polarization and a perpendicular magnetic field is detected in the inner 75-mas jet \cite[][]{AG95,dal96,jiang96,cotton97}. The inner 10 mas region contains two compact components, having comparable flux densities \cite[][]{zhang94}. From previous studies, it was unclear as to which component was associated with the core from available data, as both components were larger than typical VLBI cores of radio-loud AGN, and indicate a stronger polarization compared to regular AGN cores. The jet bends at $\sim75$~mas away from the brighter component identified as C2 in \cite{zhang94}, from a position angle P.A. of $-135^\circ$ to $-115^\circ$, where a changing polarization angle is also indicated \cite[][]{AG95,lud98}. The outermost portion of the VLBI jet aligns well with the kpc-scale jet and lobe.

In section \ref{sec:data}, the VLBI observations used for the morphology and kinematics studies are summarized and the basic data reduction procedure is reviewed. In section \ref{sec:results}, we identify the core by investigating the spectral index and compactness of VLBI components and infer the jet proper motion speed, and discuss the jet morphological characteristics at the pc and kpc scales. In section \ref{sec:discussion}, the observed VLBI jet properties and possible $\gamma$-ray emission mechanism and sites as applicable to 3C 286 are discussed, followed by a summary in section \ref{sec:summary}. A standard $\Lambda$-cold dark matter cosmological model with H$_0$ = 73 km s$^{-1}$ Mpc$^{-1}$, $\Omega_{\rm M} = 0.27$ and $\Omega_\Lambda = 0.73$ \cite[][]{Spergel07} is employed in our study. At a redshift of $z = 0.849$, 1 milliarcsecond (mas) angular size corresponds to 7.46 pc projected linear size, and the conversion from angular velocity to projected linear speed is 1 mas yr$^{-1}$ = $45\,c$.

%%%%%%%%%%%%%%%%%%%%%%%%%%%Fig1 %%%%%%%%%%%%%%%%%%%%%%%%%%
%%%%%%%%%%%%%%%%%%%%%%%%%%%/FIG %%%%%%%%%%%%%%%%%%%%%%%%%

\section{Description of VLBI data}
\label{sec:data}

The VLBI data used in this paper were obtained from the Very Long Baseline Array (VLBA) archive \footnote{https://archive.nrao.edu/archive/advquery.jsp}. Table \ref{tab:obs} summarizes the observation frequencies and epochs. These observations have diverse setups and research purposes: a 2.3/8.6-GHz Very Long Baseline Array (VLBA) project studying AGN core shift in which 3C 286 was used as a calibrator \cite[project code: RDV23, RDV37 ---][]{sok11}, the VSOP pre-launch survey at 5 GHz \cite[project code: BH019 --- ][]{fom00}, the VLBA Calibrator Survey (VCS) program at 8.3 GHz\footnote{http://gemini.gsfc.nasa.gov/vcs/} \cite[project code: BB023, BF025 ---][]{bea02,pet08}, the Monitoring Of Jets in Active galactic nuclei with VLBA Experiments (MOJAVE)\footnote{http://www.physics.purdue.edu/astro/MOJAVE/index.html} \cite[project code: BK016, BK048, BL129 ---][]{Lister09}, and a VLBA water maser observation and two astrometric observations of wobbling jets of blazars in which 3C 286 was used as a calibrator at 22~GHz (BS084, BA082, and BA084). The data at various frequencies reveal structures at diverse size scales, as can be seen in Figure \ref{fig:mor}.

The primary data reduction procedure including editing, amplitude calibration, ionosphere correction, and fringe fitting have already been performed using the astronomical image processing system (AIPS), developed by the National Radio Astronomy Observatory (NRAO), USA. We only performed a few iterations of self-calibration in Difmap \cite{she94} to calibrate the residual phase and amplitude errors. After self-calibration, the ($uv$) visibility data were fitted with several Gaussian components in Difmap using the {\rm MODELFIT} program.

\textsc{\begin{table}
%\renewcommand{\baselinestretch}{.4}
%\small
%\centering
\caption{Description of the archival VLBI observations used in this paper.} 
\scalebox{.85}{
  \begin{tabular}{cccc} \hline
Code  & Frequency (GHz) & Epoch            & Bandwidth (MHz) \\ \hline
   BK016  & 15.3   & 1995 April 9     & 32       \\
   BB023  & 8.3    & 1996 May   15    & 32         \\
   BH019  & 4.9    & 1996 June 06     & 64       \\
   BF025  & 8.3    & 1997 January 11  & 16       \\
   BK048  & 15.3   & 1997 March 10    & 60       \\
   RDV23  & 8.6    & 2000 October 23  & 32         \\
   BS084  & 22.2   & 2002 July 1      & 32          \\
   BR077  & 15.3   & 2002 August 12   & 60       \\
   RDV37  & 8.6    & 2003 March 10    & 32       \\ 
   BL129  & 15.3   & 2005 June 22     & 64         \\
   BA082  & 22.2   & 2006 July 17     & 32       \\
   BK134  & 1.4    & 2007 April 30    & 32      \\
   BA084  & 22.2   & 2009 December 16 & 32        \\\hline
  \end{tabular}}
\label{tab:obs}
\end{table}}

\begin{table*}
%\centering
%\renewcommand{\baselinestretch}{1.5}
%\small
\caption{Image parameters in Figure 1.}
\begin{tabular}{lclcccc}
\hline \hline
Figure  & Epoch &  Band  &  $S_{\rm peak}$ & Contours     & Beam FWHM and P.A. \\
Label   & (yr)  & (GHz)  &  (Jy beam$^{-1}$) & (mJy beam$^{-1}$)& (mas$\times$mas, $^\circ$)\\
\hline
 Fig. \ref{fig:mor}a&  2007.33&  1.4&  1.71   & 6.0$\times$($-$1,1,2,$\ldots$,64)& 8.1$\times$5.4, $-$15.9 \\
 Fig. \ref{fig:mor}b&  2007.33&  1.4&  1.13   & 6.0$\times$($-$1,1,2,$\ldots$,64)& 6.7$\times$2.4, $-$15.0 \\
 Fig. \ref{fig:mor}c&  1996.43&  4.9&  0.41   & 6.0$\times$($-$1,1,2,$\ldots$,64)& 2.6$\times$1.6, 0 \\
 Fig. \ref{fig:mor}d&  1997.03&  8.3&  0.27   &15.0$\times$($-$1,1,2,$\ldots$,16)& 2.6$\times$1.6, 0\\
 Fig. \ref{fig:mor}e&  1995.27& 15.3&  0.24   & 5.0$\times$($-$1,1,2,$\ldots$,32)& 2.6$\times$1.6, 0\\
 Fig. \ref{fig:mor}f&  2006.54& 22.2&  0.17   & 4.0$\times$($-$1,1,2,$\ldots$,32)& 0.8$\times$0.5, 0\\ 
\hline
\end{tabular}
%\\[0.1cm]
\label{tab:map}
\end{table*}

\section{Results}
\label{sec:results}

\begin{figure*}
\begin{center}
\includegraphics[width=0.38\textwidth]{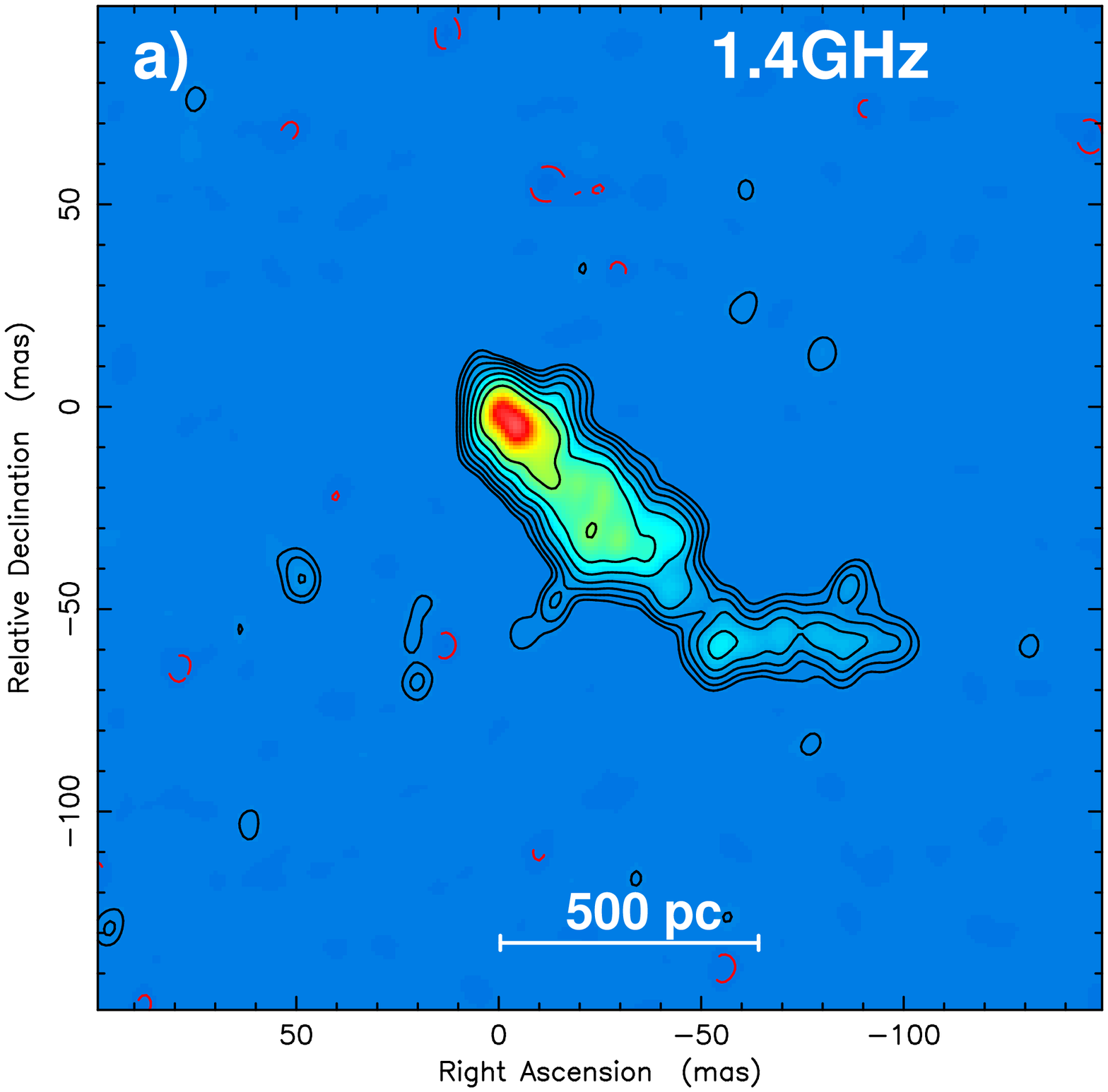} \hspace{5mm}
\includegraphics[width=0.38\textwidth]{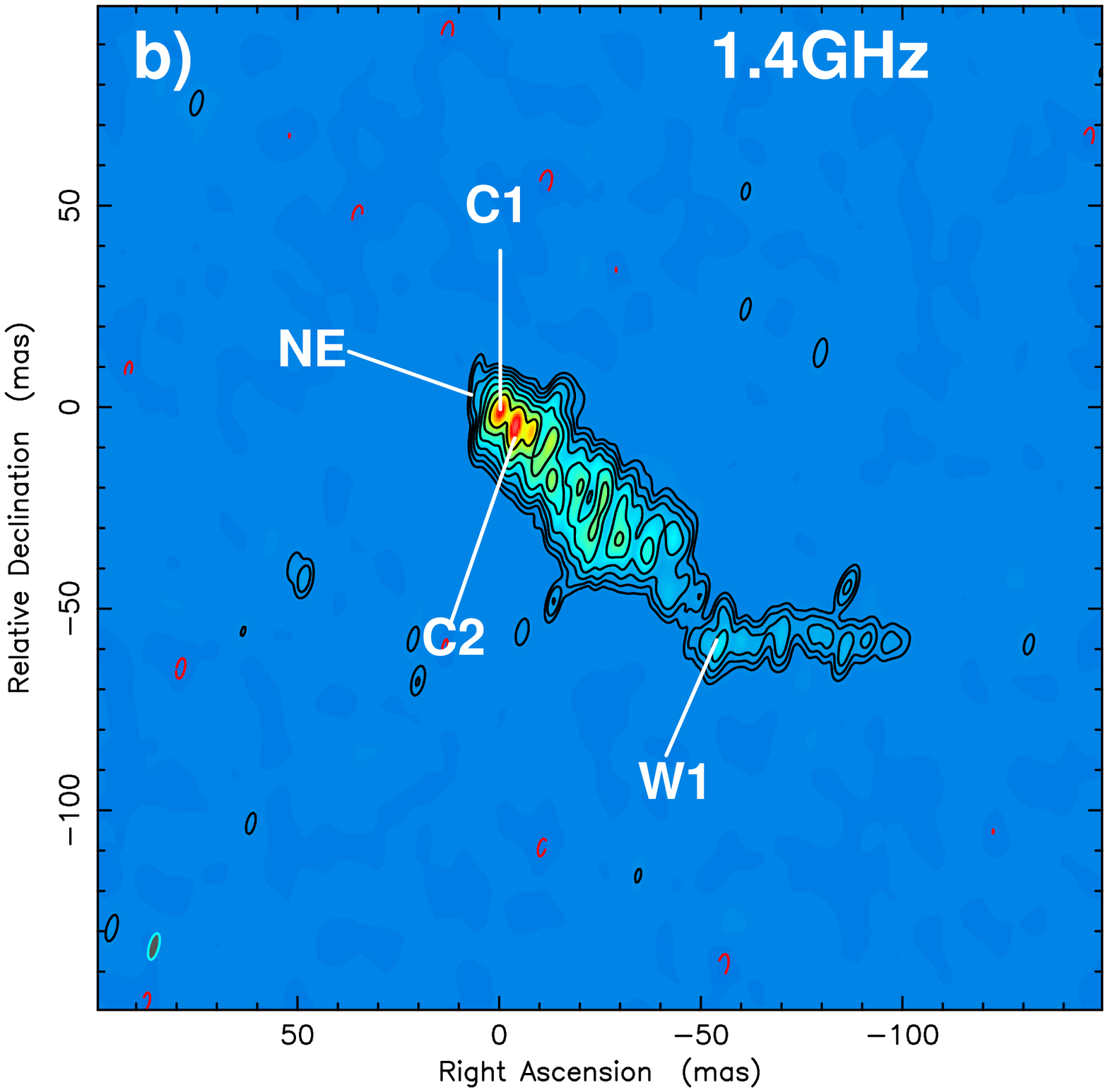} \hspace{5mm} 
\includegraphics[width=0.38\textwidth]{fig2c.eps} \hspace{5mm} 
\includegraphics[width=0.38\textwidth]{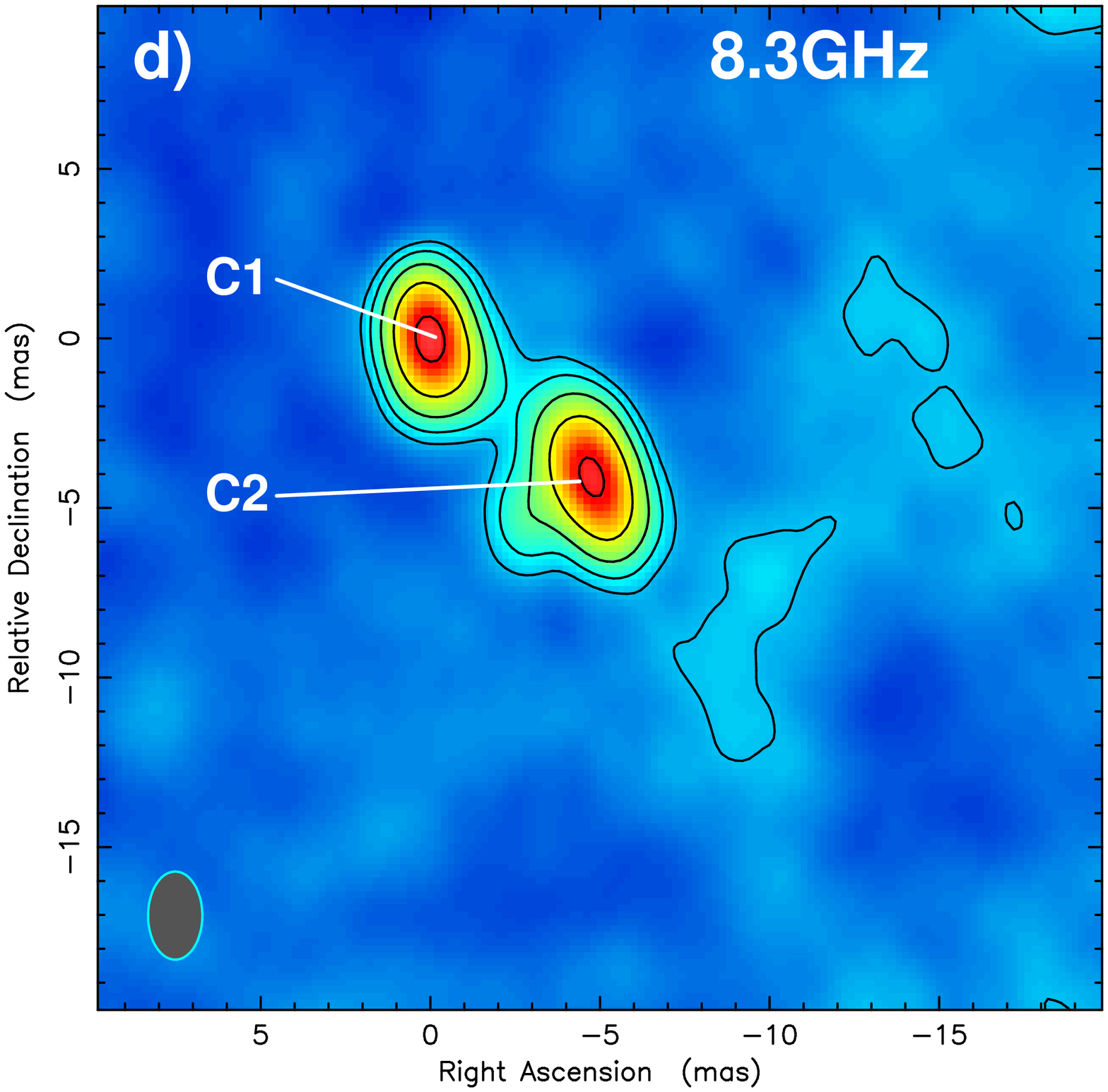} \hspace{5mm}
\includegraphics[width=0.38\textwidth]{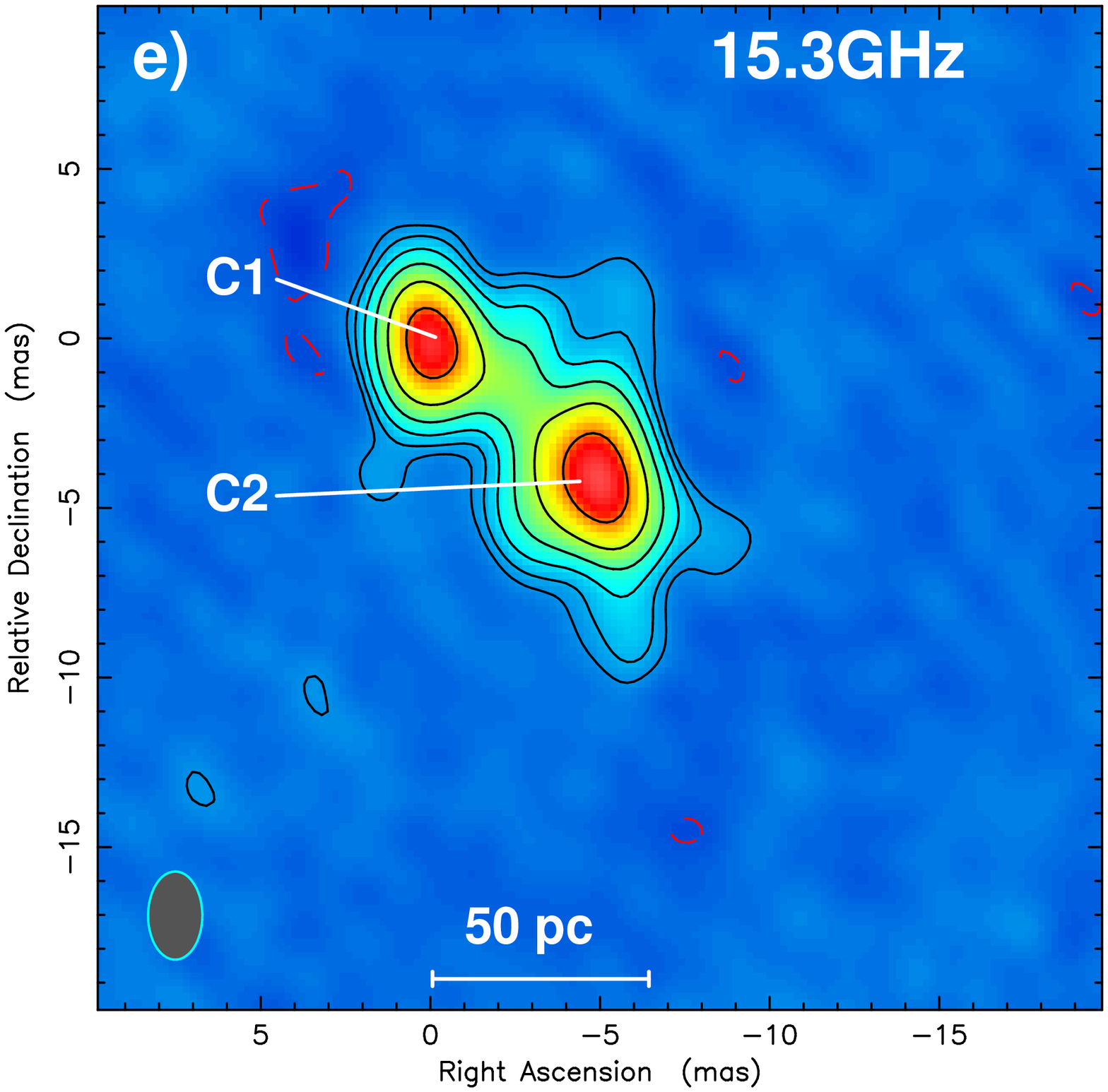} \hspace{5mm}
\includegraphics[width=0.38\textwidth]{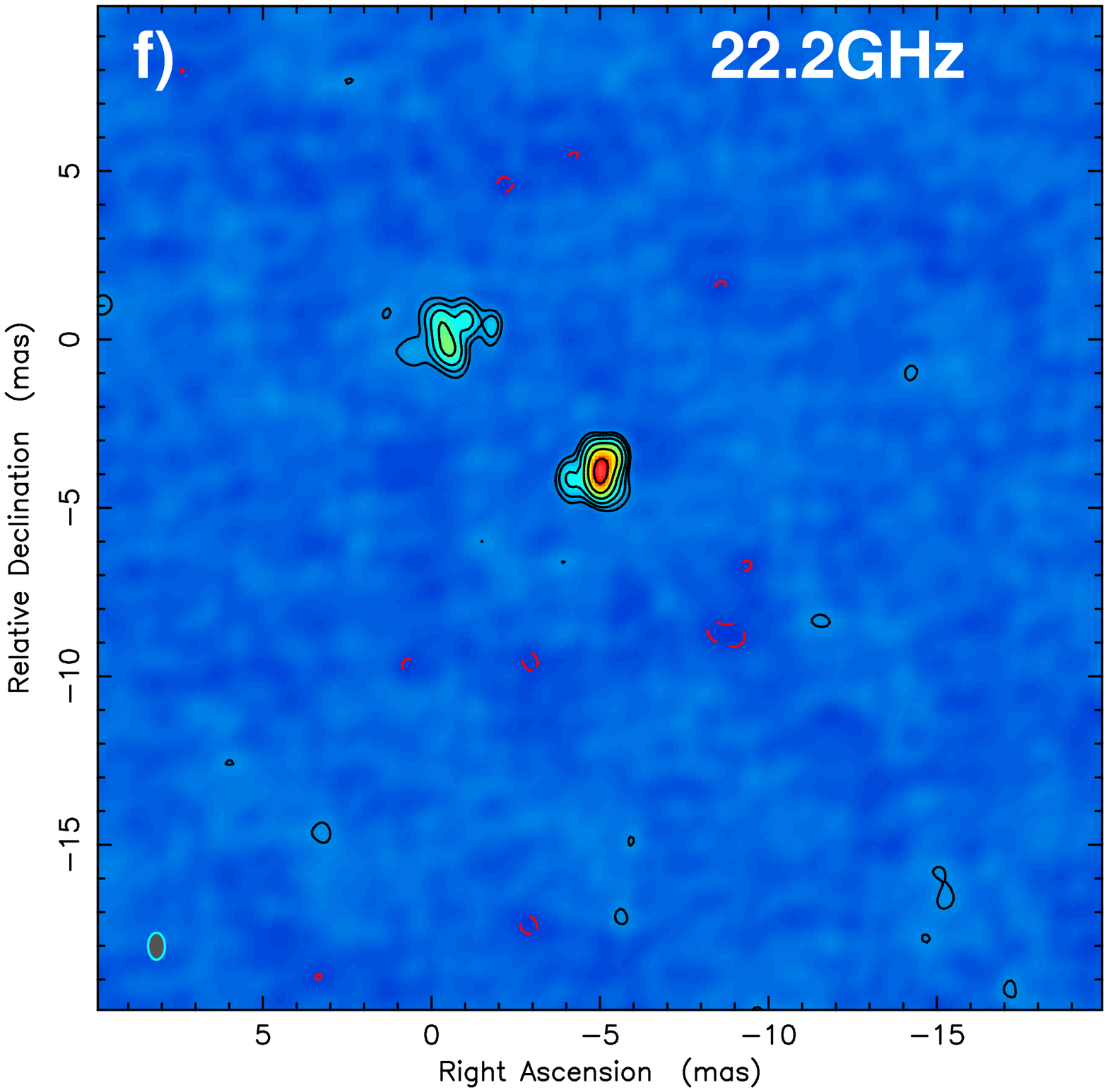} 
\caption{Radio emission structure of 3C~286 on pc scales.
Panel {\it a}: the natural-weighting image; panel {\it b}: the uniform-weighting image. The ellipse at the left-bottom corner in each panel shows the restoring beam.
The image parameters are listed in Table \ref{tab:map}. The image centers have been shifted to the position of the core (C1).
}
\label{fig:mor}
\end{center}
\end{figure*}

Representative images at five frequencies, 1.4 GHz, 4.9 GHz, 8.3 GHz, 15.3 GHz and 22.2 GHz, are presented in Figure \ref{fig:mor}. In order to identify and compare pc-scale emission structures with respect to emitted flux density level (flux density variations and spectral index), kinematic properties and their evolution with observational epoch, the images at 4.9, 8.3 and 15.3 GHz have been restored with the same beam size of 2.6 mas $\times$ 1.6 mas (full width at half maximum, FWHM). The 1.4 GHz images, restored with beam sizes of 8.07 mas $\times$ 5.35 mas and a higher resolution 6.7 mas $\times$ 2.4 mas, respectively are instructive in the identification and study of jet morphological structures at larger pc--kpc scales and are used to compare our results with previous studies. The mapping results and contour levels for all observations are summarized in Table \ref{tab:map} and results from Gaussian model fitting of clearly discernible components C1 and C2 are presented in Table \ref{tab:mod}. The fit parameter uncertainties were estimated considering model errors and using the expressions presented in \citep{fom99}, based on the post-fit root-mean-square error associated with the image pixels. The uncertainties in flux density are assumed to be 10 \% and the errors in position are assumed to be $\sim 1/5$ of the beam size. We now discuss inferences that can be drawn from these images, and the flux densities and kinematics from the component fitting.

The jet morphology at pc--kpc scales obtained from the 1.4 GHz image is presented in Figure \ref{fig:mor}-a which is constructed with a natural weighting, showing a narrow ribbon-like structure continuously extending to the southwest with a total extent of $\sim$120 mas, corresponding to a projected linear size of 910 pc. The bending of the jet from P.A. of $-135^\circ$ to $-115^\circ$ is evident at a distance of about 80 mas (600 pc). The overall morphology is consistent with that obtained in 1991 by \cite[][]{zhang94}. The integrated flux density in Figure \ref{fig:mor}-a is 11.26 Jy, 20\% lower than in the 1991 image, possibly attributable to long-term variability (duration $\sim$ 2 decades) or only a fraction of extended emission being resolved in the current data. The brightest region at the northeastern end of the continuous emission structure appears marginally resolved at a resolution of 8.07 mas $\times$ 5.35 mas. The higher-resolution image in Figure \ref{fig:mor}-b is constructed with the same restoring beam as that used for the 1991 image, i.e. 6.7 mas $\times$ 2.4 mas, and indicates a good agreement with that presented by \cite{zhang94} which was obtained with a full-track global VLBI array. Minor differences in appearance of the images arise from the fact that the current image was made from snapshot data resulting in the smooth underlying jet stream being resolved to an extent. The image indicates that the main jet body is resolved into a series of knots. The bright base is clearly resolved into two components, C1 and C2 \cite[here, we use the same labels as][]{zhang94}, whose peak intensities are inferred to be 0.91 Jy beam$^{-1}$ and 1.13 Jy beam$^{-1}$ respectively, slightly brighter than 16 years earlier. The brightest components C1 and C2 are the most probable locations of the core. A northeastern tip extends out from component C1 which could be associated with a counterjet. Figure \ref{fig:mor}-c at 4.9 GHz reveals the detailed structure of the inner 20 mas (150 pc) jet. This image is also characterized by knotty features similar to the 1.4-GHz image. The components C1 and C2 are detected at a separation of $\sim$6 mas along P.A.$=-132^\circ$, in agreement with the imaging results at the same frequency in the literature including the global VLBI array observations reported by \cite{zhang94}, the European VLBI Network (EVN) \citep{jiang96} and the VLBA \citep{cotton97}. Since the current data were compiled in snapshot observations and owing to the lack of short ({\it uv}) spacings, the weak and extended jet emission beyond 20 mas is not well sampled. The flux densities of C1 and C2 are 0.38 Jy and 1.55 Jy, respectively, and the deconvolved sizes of C1 and C2 are 2.7 mas and 3.7 mas. Compared to the flux densities reported in the early 1990s, C2 remains the same, but C1 appears to have weakened. Figures \ref{fig:mor}-d and \ref{fig:mor}-e show the morphologies at 8.3 and 15.3 GHz, respectively. At these two higher frequencies, only the two bright and compact components C1 and C2 are detected. The diffuse jet downstream disappears owing to its steep-spectrum nature, the higher frequency and high resolution of the images. The peak intensities of C1 and C2 are the same at both frequencies, while the integrated flux density of C1 is lower than C2 as C1 is more compact. Figure \ref{fig:mor}-f shows the 22.2 GHz image in which only two compact knots are detected with C2 being $\sim$ 1.5 times brighter than C1. 

%%%%%%%%%%%%%%%%%%%%%%%%%%%Fig2 %%%%%%%%%%%%%%%%%%%%%%%%%%
%%%%%%%%%%%%%%%%%%%%%%%%%%%/FIG %%%%%%%%%%%%%%%%%%%%%%%%%

\section{Discussion}
\label{sec:discussion}

\begin{figure}
\centerline{\includegraphics[width=0.5\textwidth]{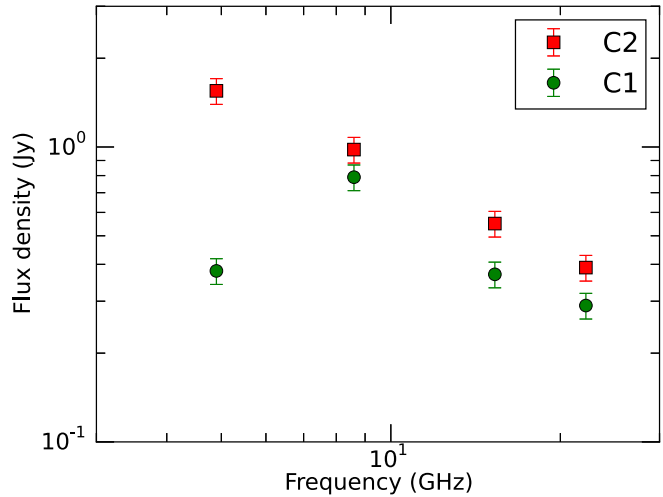}} 
\caption{Flux density as a function of observing frequency for components C1 and C2 in the VLBI images.}
\label{fig:VLBIspec}
\end{figure}

The flux densities of components C1 and C2 as a function of observation frequency are presented in Figure \ref{fig:VLBIspec}. Considering that the VLBI features show moderate-level variability at high frequency (see Table \ref{tab:mod}), we only included data points at close epochs from available archival data.
The component C2 shows a steep spectrum with a spectral index of $\alpha = -1.4$. This is a typical value for an optically thin jet, indicating that C2 is a jet knot. The component C1 shows an inverted spectrum with a turnover between 5 and 8 GHz, implying an absorption (either synchrotron self-absorption or free-free absorption). Based on the spectral properties, albeit derived from non-simultaneous observations, and the moderate-level variability, we may infer that the core is associated with the component C1. Besides, C1 is more compact than C2 at all epochs and frequencies, as is evident from Table \ref{tab:mod}, providing additional support to its identification as the core. The use of higher-frequency and higher-resolution data thus provides confidence in the evidence for the core identification.
The brightness temperature of C1 ranges from $0.7\times 10^9 $ to $9.0 \times 10^9 $ K, with a mean value of $2.8 \times 10^9$ K. This core brightness temperature is remarkably lower than that for the flat-spectrum radio-loud AGN which have typical core brightness temperatures ranging from $10^{11-13}$~K  \cite[e.g.][]{kov05}. The relatively lower brightness temperature of C1, compared to typical radio-loud AGN cores, may indicate that the core is likely beamed away from the observer (also see the illustration in \cite{cotton97}).
%emission from C1 includes both the genuine (self-absorbed) core and some optically-thin jet emission. 

\begin{figure}
\centerline{\includegraphics[width=0.5\textwidth]{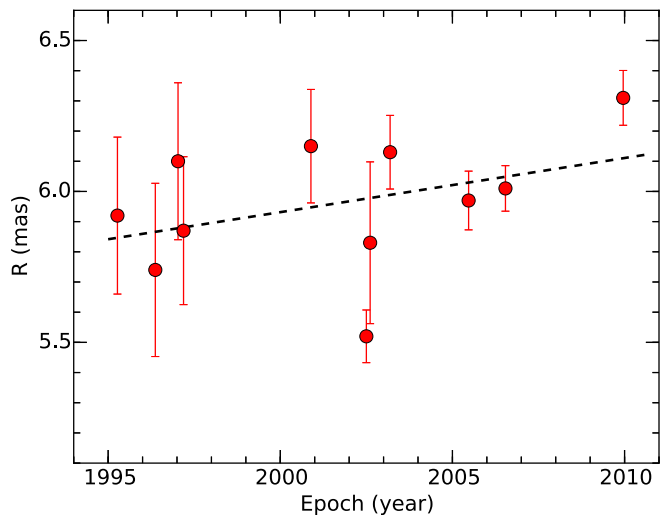}}
\caption{Proper motion of the jet component C2.
The dashed line represents the least-squares linear fit.
The slope of the fit corresponds to the proper motion of $0.013 \pm 0.011$ mas yr$^{-1}$.
}
\label{fig:pm}
\end{figure}

Figure \ref{fig:pm} shows the time evolution of the separation between C2 and C1. The dashed line is a linear fit to the data from which we infer a proper motion (slope) of $0.013 \pm 0.011$ mas yr$^{-1}$, corresponding to a linear speed $\beta_{\rm app} = 0.6 \pm 0.5$.
The fit indicates a large uncertainty, which in fact reflects the heterogeneity of the observational data: inhomogeneous choices of observation frequencies, different resolutions and {\it uv} coverage. In addition, the small number of the data points could also contribute to the large statistical error of the sparsely distributed data points. Several measures have been made to minimize this heterogeneity, including the use of only 8.3, 15.3 and 22.2 GHz data, which have similar resolution; the setting of a common lower limit of the {\it uv} range, with the deletion of visibilities below it so that model fitting is less contaminated by extended emission; and least-square fitting by using the same-frequency data (though they are fewer), such that the derived results from data in the above three frequency bands are consistent.
With the available data and accounting for all the mentioned sources of errors, the resulting proper motion is the best estimate for this source. Additional data points from future observations can improve the accuracy of this estimate. 

As discussed above, if C1 is the core, then $\beta_{\rm app} = 0.6$ is the apparent jet advancing speed. For two-sided jets, Doppler boosting enhances the apparent brightness of the advancing jet, and dims the receding one.  The flux densities of C2 and the NE components indicate a clear demarcation, possibly as the NE component is receding away and is hence, Doppler de-boosted. In addition, it is inferred that NE and C2 are at similar distances from C1 indicating that NE can be identified as the counterjet. The jet-to-counterjet intensity ratio \cite[e.g.][]{1985ApJ...295..358L} is given by,
\begin{equation}
J = \frac{S_{\rm j}}{S_{\rm cj}} = \left( \frac{1+\beta\cos\theta}{1-\beta\cos\theta} \right)^{n+\alpha},
\end{equation}
where $\beta$ is the true jet speed, $\theta$ is the inclination angle towards the observer's line of sight, $S_{\rm j}$ is the flux density in the approaching jet, $S_{\rm cj}$ is the flux density of the counterjet and the index $n = 3$ for a resolved jet knot and $n = 2$ for a continuous flow. From the measured flux density of the counterjet component NE ($\sim$ 0.07 Jy) and the jet component C2 (1.55 Jy) that are at similar, opposite distances from the core, we obtain $J = 23$. Using the measured $\alpha = 1.4$ (for the steep spectrum component C2) and $n = 3$, we get $\beta\cos\theta = 0.34$. From the apparent proper motion speed 
\begin{equation}
\beta_{\rm app} = \frac{\beta\sin\theta}{1-\beta\cos\theta},
\end{equation}
the inclination angle $\theta = 48^\circ$, the jet speed $\beta = 0.5$ and the bulk flow Lorentz factor is $\sim$1.2. Taking the upper limit of the apparent speed $\beta = 1.1$ into the calculation, we get $\theta = 65^\circ$, $\beta = 0.8$ and Lorentz factor $\sim 1.7$. The 3C 286 jet is among the fastest in the high-power compact symmetric objects (CSOs) and CSSs, consistent with the conclusion that a CSO with a faster jet tends to be brighter in radio \cite[][]{An12,AB12}. On the other hand, the inferred jet speed and Lorentz factor are significantly lower than that in $\gamma$-ray blazars, which generally host highly relativistic jets, often displaying superluminal motion \cite[e.g.][]{jor01}. This implies that the 3C 286 jet could be intrinsically different from the blazar jets, possibly in the launching region or acceleration zone properties or in the jet composition.

Most {\it Fermi}-detected AGN are characterized by strong relativistic beaming. The observed properties of the majority of $\gamma$-ray AGN include compact cores, high core brightness temperatures, rapid and prominent variability, high Lorentz factor and apparent superluminal jet speed \cite[e.g.][]{homan06,hov09,lis11}. The $\gamma$-ray emission from blazars is generally interpreted by a single-zone jet model in which $\gamma$-rays are produced in the innermost relativistic jet via synchrotron self-Compton process \cite[e.g.][]{mar92} or by the Compton up-scattering of external seed photons from the disk and torus, as explained earlier. However, unlike blazars, the misaligned sources, i.e., the non-blazar AGN are not strongly beamed and their jet is not extremely relativistic. Thus, models for production of $\gamma$-ray emission from blazars may not be applicable for CSO and CSS sources, necessitating the exploration of other possibilities. For example, extended $\gamma$-ray emission was also detected from Cen A lobes \cite{Abdo10-CenAlobe}, which is interpreted as the inverse Compton scattering of the cosmic microwave background photons by the relativistic electrons injected by the jets into the lobes. Although there are some predictions of high-energy emission from compact radio sources in hard X-ray and $\gamma$-ray energy bands, it is still unclear why there is a deficiency of misaligned compact radio sources (CSOs, CSSs) among {\it Fermi}/LAT-detected AGN as theoretical calculations indicate that $\gamma$-ray emission is expected from CSO and CSS \cite[e.g.][]{sta08,kino09a,kino09b,mig14}. In these scenarios, the relativistic electrons accelerated in and injected from the hot spots to the CSO lobes can up-scatter the ultraviolet photons originating from the accretion disk to GeV energy range \cite[e.g.][]{sta08}. Alternatively, Bremsstrahlung emission from the shocked plasma created by the expanding cocoon of the young radio sources (CSO and CSS) may also significantly contribute to the high-energy emission \cite[e.g.][]{kino09a}. Until now, there are only a few identified $\gamma$-ray emitting CSO and CSS and candidates, including  4C $+$55.17 \cite[][]{mcc11}, PMN J1603$-$4904 \cite[][]{mul14}, 2234+282 \cite[][]{An16a}, PKS 1718$-$649 \cite[][]{mig16}, 0202+149 \cite[][]{An16b}, 4C $+$39.23B, and 3C 286 (the present work). The main obstacle in detecting isotropic $\gamma$-ray emission generated in the lobes or cocoons of misaligned AGN is that they require high sensitivity observations. Since the electron density and temperature are much higher in the young radio lobes than in older ones, the youngest and most compact radio galaxies are much easier to detect as $\gamma$-ray emitters \cite[][]{kino09a}. 

3C 286 has been a target of interest as a compact radio quasar in 1980s and 1990s, owing to its stable flux density enabling its use as a calibrator source. The source was detected in the 100 MeV -- 100 GeV energy range with a $\gamma$-ray flux of $4.1 \times 10^{-12}$ erg cm$^2$ s$^{-1}$ \cite[][]{ace15} in {\it Fermi}-LAT observations. As discussed above, the jet may not contribute significantly to the $\gamma$-ray emission in this source. Instead, a main contributor to the $\gamma$-ray flux could be the interaction between the mildly relativistic advancing jet and the interstellar medium which results in the shock heating of the gas at the interface resulting in the extended $\gamma$-ray emission. A propagating shock in a relativistic jet tends to transfer its kinetic energy into thermal energy, given by $E_{\rm th.} = kT \sim \frac{3}{16} m_h \beta^2 c^2$ \citep[][]{2002apa..book.....F}, where $m_h$ is the mass of the proton which constitutes the shocked gas. The heated post-shock subsonic gas temperature here corresponds to $T \sim 5 \times 10^{10}$ K for $\beta = 0.07$ at the kpc scale. This heated gas can emit MeV photons over short durations assuming that the gas density is sufficiently high and that the cooling time is appropriate to enable this. Further, recent studies focused on the detection of CSOs indicate that MeV--GeV emission can be produced also through the up-scattering of circum-nuclear infrared to ultraviolet photons by relativistic electrons in the compact, expanding radio lobes, such as in PKS 1718$-$649 \cite[e.g.][]{mig16}. A SED fit of 3C 286 from radio to $\gamma$-ray bands may offer evidence to distinguish between these models.

The studies of $\gamma$-ray emission from non-blazar AGN is particularly important for understanding the physical origin and composition of the extragalactic $\gamma$-ray background (EGB) in GeV energy bands which remains an unsolved fundamental question in astrophysics \cite[see a recent review by][]{FS15}. In spite of the observation that blazars constitute the dominant population amongst the extragalactic $\gamma$-ray sources, the contribution of the point-like blazars to the GeV diffuse $\gamma$-ray background is still not well understood \cite[][]{kne08}. Some early studies found that blazars can account for the entire EGB observed by EGRET \citep[e.g.,][]{SS96}. However, the high sensitivity of {\it Fermi}, offering an opportunity to detect fainter sources than EGRET \cite[][]{ace15}, led to the discovery of many non-blazar sources including starbursts and radio galaxies \cite[][]{Abdo10-MAGN}. Recent investigations of the composition and contribution of the $\gamma$-ray emitters, especially in the energy range $<$ 100 GeV even implied that unresolved blazars only contribute $\sim 20\%$ of the diffuse EGB \cite[e.g.,][]{muc00,Abdo10-EGB,2011ApJ...736...40S,2015ApJ...800L..27A}. According to the orientation-based unification schemes of radio-loud AGN, blazars are a sub-population with a relativistic jet pointing very close to the observer's line of sight \cite[][]{UP95}. Misaligned AGN, including the current source, 3C 286, with jets possibly beamed away from the observer's line of sight are less luminous but are more numerous than blazars. They could thus contribute significantly to the isotropic EGB at the MeV--GeV energies.

%%%%%%%%%%%%%%%%%%%%%%%%%%%Fig 2%%%%%%%%%%%%%%%%%%%%%%%%%%
%%%%%%%%%%%%%%%%%%%%%%%%%%%/FIG 2%%%%%%%%%%%%%%%%%%%%%%%%%

%%%%%%%%%%%%%%%%%%%%%%%%%%%Fig 3%%%%%%%%%%%%%%%%%%%%%%%%%%
%%%%%%%%%%%%%%%%%%%%%%%%%%%FIG 3%%%%%%%%%%%%%%%%%%%%%%%%%

\section{Summary}
\label{sec:summary}

The pc-scale radio properties of the $\gamma$-ray quasar 3C 286 are investigated in detail. The radio core is for the first time identified from its compact structure detected in high-frequency, high-resolution VLBI images and its inverted radio spectrum.
With the available data, we obtained a jet proper motion of $0.6\pm0.5\, c$, resulting in a jet speed of $\sim 0.5\,c$ and an inclination angle of $\sim 48^\circ$. Large proper motion values are excluded by the data. Improving the accuracy requires more VLBI observations at the same frequency (e.g., 15 GHz) over a sufficiently long duration. This preliminary proper motion measurement implies a mildly relativistic, subluminal jet moving at a moderate inclination angle.   
Considering the low Lorentz factor and low beaming effect of the jet, $\gamma$-ray emission from 3C 286 and other non-beamed compact AGN may have different origin from blazars where the $\gamma$-ray emission mainly originates from the inner beamed relativistic jet. Multi-band SED fitting may provide complementary support to this argument. Other physical mechanisms must be considered to discern the production sites of $\gamma$-ray emission in the misaligned compact AGN population, including expanding lobes and cocoons, and the shocks created from the interaction of the jet with the surrounding ambient medium. 
Although the sample size of observed $\gamma$-ray CSO and CSS sources is still small due to the insufficient sensitivity of $\gamma$-ray telescopes, these compact misaligned sources constitute a very interesting class of $\gamma$-ray emitters. Owing to the effectively larger emission surface area and the expected number of events, these compact young radio sources are expected to contribute in part to the diffuse extragalactic $\gamma$-ray background.

\begin{table*}
%\centering
%\renewcommand{\baselinestretch}{1.5}
%\small
  \begin{tabular}{cccccccc} \hline
Epoch   & Freq & Comp & $S$             & $R$    &P.A.      &$\theta$ & $T_{\rm b}$ \\
        &(GHz) &      &(Jy)           &(mas) &($^\circ$) &(mas)    & ($10^{9}$K) \\
 (1)    &(2)   &(3)   &(4)            &(5)   &(6)     &(7)      &(8)          \\\hline
1995.27 &15.3  & C1    & 0.37$\pm$0.04 &  0    & 0      &1.6     &  1.5 \\
        &      & C2    & 0.55$\pm$0.05 & 5.92  &$-$130.2&2.5     &      \\
1996.37 &8.3   & C1    & 0.18$\pm$0.02 &0      &0       &1.9     &  2.4    \\
        &      & C2    & 0.16$\pm$0.02 &5.74   &$-$136.7&2.3     &      \\
1996.43 &4.9   & C1    & 0.38$\pm$0.04 &  0    & 0      &2.7     &  4.8     \\
        &      & C2    & 1.55$\pm$0.16 & 6.58  &$-$126.8&3.7     &      \\
1997.03 &8.3   & C1    & 0.42$\pm$0.04 &  0    & 0      &1.6     &  5.3 \\
        &      & C2    & 0.50$\pm$0.05 & 6.10  &$-$131.8&2.1     &      \\
1997.19 &15.3  & C1    & 0.25$\pm$0.03 &  0    & 0      &1.5     &  1.0     \\
        &      & C2    & 0.31$\pm$0.03 & 5.87  &$-$130.0&2.2     &  \\
2000.89 & 8.6  & C1    & 0.52$\pm$0.05 &0      &0       &1.6     & 9.0 \\
        &      & C2    & 0.12$\pm$0.01 &6.15   &$-$126.9&1.9     &  \\
2002.50 &22.2  & C1    & 0.29$\pm$0.04 &0      &0       &1.2     & 0.5 \\
        &      & C2    & 0.39$\pm$0.04 &5.52   &$-$132.5&0.3     &   \\  
2002.61 &15.3  & C1    & 0.37$\pm$0.04 &  0    & 0      &2.0     &  0.9   \\
        &      & C2    & 0.55$\pm$0.06 & 5.83  &$-$128.8&2.9     &    \\
2003.19 &8.6   & C1    & 0.79$\pm$0.08 &  0    & 0      &3.0     &  2.6     \\
        &      & C2    & 0.98$\pm$0.10 & 6.13  &$-$129.0&3.3     &   \\  
2005.47 &15.3  & C1    & 1.04$\pm$0.10 &0      &0       &2.1     & 3.2\\
        &      & C2    & 1.20$\pm$0.12 &5.97   &$-$130.5&1.9     &\\
2006.54 &22.2  & C1    & 0.13$\pm$0.01 &0      &0       &1.1     & 0.7\\
        &      & C2    & 0.29$\pm$0.03 &6.01   &$-$131.3&0.5     &\\
2009.96 &22.2  & C1    & 0.14$\pm$0.01 &0      &0       &1.1     &0.8\\
        &      & C2    & 0.23$\pm$0.02 &6.31   &$-$127.4&1.5     &\\\hline \\
\end{tabular}
\label{tab:mod}
\vspace{0.1cm}
\caption{Parameters of the fitted Gaussian model components and the inferred brightness temperatures.
Columns in this table are as follows: 
Col. 1, observation epoch;
Col. 2, observing frequency;
Col. 3, component label;
Col. 4, integrated flux density of each component;
Col. 5 and Col. 6, distance and position angle of each component with respect to component C1;
Col. 7, size of the deconvolved Gaussian model,
Col. 8, brightness temperature of the core.}
\end{table*}

\section*{Acknowledgements}
We thank the referee for helpful comments.
The authors are supported in this work by SKA pre-construction funding from the China Ministry of Science and Technology (MOST) and the Chinese Academy of Sciences (CAS), the China--Hungary Collaboration and Exchange Programme by the International Cooperation Bureau of the CAS.
TA thanks the grant support by the Youth Innovation Promotion Association of CAS. 
PM thanks the grant support by the CAS President's International Fellowship for Postdoctoral Researchers program and the National Natural Science Foundation (grant no. 11650110438). 
ZLZ thanks the One Hundred Talents Programme of the CAS.
We thanks Sandor Frey for helpful comments.
This research has made use of data from the MOJAVE database that is maintained by the MOJAVE team \cite[][]{Lister09}. 
The National Radio Astronomy Observatory is a facility of the National Science Foundation operated under cooperative agreement by Associated Universities, Inc. 
This work has made use of NASA Astrophysics Data System Abstract Service, and the NASA/IPAC Extragalactic Database (NED) which is operated by the Jet Propulsion Laboratory, California Institute of Technology, under contract with the National Aeronautics and Space Administration.

%%% references

\end{document}